\documentclass[%
superscriptaddress,
preprint,
showpacs,preprintnumbers,
amsmath,amssymb,
aps,
prl,
]{revtex4-1}

\usepackage{color}
\usepackage{graphicx}
\usepackage{dcolumn}
\usepackage{bm}

\newcommand{\R}{{\mathbb R}}
\newcommand{\Z}{{\mathbb Z}}

\begin {document}
\preprint{APS/123-QED}

\title{Distributional Response to Biases in Deterministic Superdiffusion}

\author{Takuma Akimoto}
\email{akimoto@z8.keio.jp}
\affiliation{%
  Department of Mechanical Engineering, Keio University, Yokohama, 223-8522, Japan
}%


\date{\today}

\begin{abstract}
We report on a novel response to biases in deterministic superdiffusion.  For its reduced map, 
we show using infinite ergodic theory that the time-averaged velocity (TAV) is intrinsically random and its distribution 
obeys the generalized arc-sine distribution. A distributional limit theorem indicates that the TAV response to a bias appears 
in the distribution, which is an example of what we term a {\it distributional response} induced by a bias. 
Although this response in single trajectories is intrinsically random,  
the ensemble-averaged TAV response is linear.
\end{abstract}

\maketitle


{\it Introduction}.---Intrinsic randomness in macroscopic observables has been found in a broad range of processes more recently from 
biological transport to fluorescence of 
single nanocrystals \cite{Golding2006, Szymanski2009,Brokmann2003}. Diffusion coefficients  in biological transports show large 
fluctuations \cite{Golding2006, Szymanski2009}. The ratio of an on-state in the fluorescence of nanocrystals does not converge 
to a constant and is different in each quantum dot \cite{Brokmann2003}. The randomness of time-averaged observables can be 
characterized by power-law trapping time distributions in stochastic models, such as those in continuous time random walks (CTRWs) 
\cite{He2008, Miyaguchi2011} and dichotomous stochastic processes \cite{Margolin2006}.  Such randomness  is due to 
the breakdown of the law of large numbers stemming from a diverging mean trapping time. 
\par
Dynamical systems with infinite-invariant measures can be viewed as stochastic processes generating random time-averaged observables. 
Infinite ergodic theory plays an important role in elucidating such observables 
\cite{Aaronson1997, Thaler1998, Thaler2002, TZ2006, Akimoto2008}. It guarantees that a time average 
 of an observation function converges in distribution. In other words, time-averaged observables are intrinsically random if
 the invariant measure cannot be normalized.  In dynamical systems generating  subdiffusion, the distributional limit theorem for 
 the diffusion coefficients obtained by the time-averaged mean square displacements (TAMSDs) has been shown using infinite 
 ergodic theory \cite{Akimoto2010}.
\par
 In anomalous diffusion, the mean square displacement (MSD) grows non-linearly with time, 
 $\langle x(t)^2 \rangle \propto t^{\alpha}$ $(\alpha\ne 1)$. Diffusion is called subdiffusion if $\alpha<1$ and superdiffusion 
 if $\alpha>1$. One mechanism generating subdiffusion is the divergence in the mean trapping time describing random walks of CTRW. 
 If the mean trapping time diverges, the diffusion coefficients obtained from TAMSDs become random 
 \cite{He2008, Miyaguchi2011, Akimoto2010}. Three different mechanisms underlying superdiffusion have been identified: One stems from
  positive correlations in random walks, modeled by a fractional Brownian motion \cite{Mandelbrot1968}; The second from a persistent 
  motions in random walks, called L\'{e}vy walks \cite{Scher1975}; The third from very long jumps in random walks, called L\'{e}vy flights \cite{Ott1990}. 
  In L\'{e}vy walks and flights, the second moment of the length of persistence motion and of jumps diverges because of a power law. 
 Such power laws are  observed in Hamiltonian systems \cite{Venegeroles2009}, rotating flow \cite{Solomon1993}, 
 polymer diffusion \cite{Ott1990}, biological transport \cite{Bruno2009}, intermittent search \cite{Benichou2011}, 
 and light diffusion \cite{Barthelemy2008}. Persistent times 
 in L\'{e}vy walks and trapping times in CTRWs are characterized by indifferent fixed points in deterministic models \cite{Zumofen1993}.
Therefore, random transport coefficients will be observed in L\'{e}vy walk with the divergent mean persistent time. 
\par
 
Although random transport coefficients are universal in subdiffusion because of power-law trapping times, it is not clear whether 
time-averaged observables are intrinsically random in superdiffusions. Moreover, little is known about responses of 
time-averaged drifts (TADs) to biases in superdiffusions whereas in anomalous diffusion a generalized Einstein relation holds \cite{bouchaud90}.
In this Letter, we show a distributional limit theorem for TADs using deterministic superdiffusion models related to L\'{e}vy walks 
under bias and no bias. Surprisingly, a TAD is intrinsically random whether biased and unbiased. The result leads to a distributional response 
to bias, {\it i.e.}, the response to a bias would be characterized by a change in distribution.

{\it Model}.---
Anomalous superdiffusion that originates from a persistent motion or a long jump has been studied for chaotic dynamical systems 
\cite{Geisel1985, Miyaguchi2006}. To study a response to a bias in deterministic superdiffusions related to L\'{e}vy walks,
we propose an asymmetric deterministic diffusion model, constructed by introducing an asymmetry in the Geisel model \cite{Geisel1985}.
In particular, we consider the following map $T:\R \rightarrow \R$,
\begin{equation}
x_{n+1}=T(x_n),
\end{equation}
which have a translational symmetry $(L=\Z)$
\begin{equation}
T(x+L)=T(x)+L,
\end{equation}
and the map $T(x)$ is given by
\begin{equation}
T(x)=\left\{
\begin{array}{ll}
 (x-L)+\left(\frac{x-L}{c}\right)^z+L-1, &
 x\in [L,L+c),\\
\noalign{\vskip0.2cm}
  (x-L)-\left(\frac{L-x}{1-c}\right)^z+L+1, & x\in [L-c,L),
\end{array} \right.
\end{equation}
where $c$ is a parameter characterizing an asymmetry. This model corresponds to a L\'{e}vy walk where leftward and rightward persistent time
distributions have the same scaling exponent, {\it i.e.}, $\psi(t) \propto t^{-\beta}$ but the probabilities of the leftward and rightward walk are different.

{\it Reduced map}.---
As can be seen in Fig. 1, points near the fixed points, $x=0$ and $x=1$, on $[0,1]$ move to the left neighboring cell or the right neighboring cell,
 respectively. By translational symmetry, we can reduce an orbit of the map $T(x)$ to that of an intermittent map on $[0,1]$ (Fig. 1). For example, 
 we can obtain the following reduced map,
\begin{equation}
R(x) = \left\{
\begin{array}{ll}
x+ \left(\frac{x}{c}\right)^z, \quad {\rm mod}~1 \quad & x\leq c,\\
\\
x- \left(\frac{1-x}{1-c}\right)^z ,  \quad {\rm mod}~1\quad & x> c,
\end{array}
\right.
\end{equation}
with $0<c<1$. The invariant density of the reduced map is given by
\begin{equation}
\rho (x) =  h(x)x^{1-z} (1-x)^{1-z},
\end{equation}
where $h(x)$ is a continuous function satisfying $h(0)\ne 0$ and $h(1)\ne 0$ \cite{Thaler1983}. Thus, the invariant density cannot be normalized 
for $z\geq 2$. Consider the observation function 
\begin{equation}
f(x) = \left\{
\begin{array}{ll}
-1,  \quad & x\in [0,c_1),\\
0, \quad & x\in [c_1,c_2),\\
+1,  \quad & x\in [c_2,1),
\end{array}
\right.
\end{equation}
where $R(c_1)=1$ $(c_1<c)$ and $R(c_2)=0$ $(c_2>c)$. It follows that $X_n=f(x_1)+\cdots +f(x_n)$ is regarded as a one-dimensional random walk, 
where $x_n=R^n(x_0)$. The random walk $X_n$ corresponds to a L\'{e}vy walk where the persistent times distribution obeys a power law with 
exponent $\beta=z/(z-1)$.

{\it Dependence of EAMSD on ensemble}.---Ensemble-averaged MSD (EAMSD) and TAMSD are defined by
\begin{equation}
\langle x^2_m \rangle =\lim_{K\rightarrow\infty}\frac{1}{K}\sum_{k=0}^{K-1} (T^m(x^k)-x^k)^2,
\end{equation}
where $x^k$ is the $k$th initial point, and 
\begin{equation}
\label{tamsd}
\overline{\delta x^2_m}  =\lim_{n\rightarrow\infty}\frac{1}{n}\sum_{k=0}^{n-1} f_m(x_k),
\end{equation}
where $f_m(x)=(T^m(x)-x)^2$, respectively. For a finite invariant measure $(z<2)$, ergodicity holds, {\it i.e.}, EAMSD = TAMSD 
if initial points are distributed according to the invariant density of the reduced map. We note that ergodicity does not hold when the invariant 
measure cannot be normalized (infinite invariant measure) because an equilibrium ensemble cannot be reproduced. The impossibility to reproduce 
an equilibrium ensemble leads to aging \cite{Barkai2003}. In unbiased cases $(c=0.5)$, EAMSD is studied by the renewal theory and continuous 
time random walk \cite{Geisel1985, Shlesinger1985}:
\begin{equation}
\langle x^2_m \rangle_E \propto \left\{
\begin{array}{ll}
m^2, &z\geq2,\\
m^{3-1/(z-1)}, &\frac{3}{2} <z<2,\\
m \ln m, &z=\frac{3}{2},\\
m, &1<z<\frac{3}{2}.
\end{array}
\right.
\label{msd}
\end{equation}
For $c<0.5$, drifting motion arises in a direction toward the right.\par

Unlike hyperbolic maps, statistical quantities determined by ensemble averages significantly depend on an initial ensemble 
in intermittent maps. In particular, it is shown that the behavior of the correlation function and the power spectrum density 
depend on an initial ensemble \cite{Akimoto2007}. In renewal theory \cite{Cox}, there are two well-known processes, {\it i.e.},  
ordinary renewal and equilibrium renewal process. An initial ensemble corresponding to a specific renewal process 
is reproducible in dynamical systems. In particular, an initial ensemble for an equilibrium renewal process is an absolutely 
continuous invariant measure. Figure 2 shows that the EAMSD depends on an initial ensemble. If an initial ensemble of 
EAMSD has an invariant density (as in an equilibrium ensemble), all TAMSDs are equal to EAMSD (see Fig.~2).

{\it Lamperti-Thaler's generalized arcsine law}.---In dichotomous stochastic processes $\sigma_n$, 
an observable determined by the time average of an observation function $g(\sigma_n)$ is known to show random behavior 
if the mean residence time of a state diverges. In particular, the ratio of the occupation time of a state, $N_n/n$, 
does not converge to a constant, but converges in distribution, where $N_n$ is the occupation time of a state up to time $n$.
  The most classical example is the arc-sine law in coin-tossing: the distribution of the ratio of the period that a player 
  is on the positive side  converges to the arc-sine distribution \cite{Feller1968}. Lamperti showed that the distribution of 
  the ratio of the occupation time of a state converges to the generalized arc-sine distribution in general dichotomous 
  stochastic processes \cite{Lamperti1958}.\par 
In a dynamical system, the divergence of the mean residence time implies that the invariant density cannot be normalized 
\cite{Akimoto2010a}. Thaler has shown that the distribution of the time average of a characteristic function converges to 
the generalized arc-sine distribution \cite{Thaler1998, Thaler2002,TZ2006}. The generalized arc-sine law is valid for the time 
average of a non-$L^1(\mu)$ function, {\it i.e.}, $\int |f| d\mu =\infty$ \cite{Akimoto2008}. Lamperti-Thaler's generalized arc-sine 
(LTGA) law \cite{Thaler2002} states that for a map $S: [0,1]\rightarrow [0,1]$ satisfying 
(i) $S([0,c])=[0,1]$ and $S([c,1])=[0,1]$, (ii) $S'(x)>1$ on $(0,c]\cup [c,1)$; $S'(0)=S'(1)=1$, and (iii) 
$S(x)-x \sim a_0x^{p+1} (x\rightarrow 0)$ and $x-S(x) \sim a_1 (1-x)^{p+1} (x\rightarrow 1)$ with $p>1$ 
and constants $a_0,a_1>0$, and $c\in (0,1)$, the time average of the observation function $g(x)$ with $g(0)=a$ and $g(1)=b$ 
converges in distribution:
\begin{equation}
\Pr \left\{ \frac{1}{n}\sum_{k=0}^{n-1} g\circ S^k \leq t\right\}
\rightarrow \left\{
\begin{array}{ll}
G_{\alpha,\beta}\left(\frac{t-b}{a-b}\right) &(a>b)\\
\\
1-G_{\alpha,\beta}\left(\frac{t-b}{a-b}\right) &(a<b)
\end{array}
\right.
\end{equation}
where $\alpha=1/p$, 
\begin{equation}
\label{beta.LT}
\beta =\frac{S'(c_+)}{(a_0/a_1)^{1/p}S'(c_-)},
\end{equation} 
and the probability density function (PDF) is given by
\begin{equation}
G'_{\alpha,\beta}(t)=\frac{\beta \sin \alpha}{\pi}
\frac{t^{\alpha-1}(1-t)^{\alpha-1}}{\beta^2 t^{2\alpha}+2\beta t^{\alpha}(1-t)^{\alpha}\cos \pi \alpha +(1-t)^{2\alpha}}.
\end{equation}
This distribution is called the generalized arc-sine distribution, which emerges in a subdiffusive transport \cite{Lomholt2007} and 
weakly non-ergodic statistical physics \cite{Margolin2006, Rebenshtok2008}.
The mean of the occupation time for a state $x_k<c$, $\langle N_n/n\rangle$, is given by $\langle N_n/n\rangle =1/(1+\beta)$.
 The ensemble average of the time average of $g(x)$ is given by $(a+b\beta)/(1+\beta)$. 
We note that the exponents $\alpha$ and $\beta$ are determined by a behavior near the indifferent fixed points.

{\it Distributional response to a bias}.---LTGA law cannot be applied to the reduced map of the asymmetric deterministic diffusion
 model with $z\geq 2$ straightforwardly because the reduced map  does not satisfy the condition (i). However, the condition (i) 
 is not crucial  because an important point in LTGA law is the reinjection to the indifferent fixed points. In fact, the conditions 
 (i), (ii), and (iii) in \cite{Thaler2002} was generalized in \cite{TZ2006}. The reinjection to the fixed point $x=0$ is determined by 
 $\lim_{x\rightarrow c_1+0}R'(x)$ and $\lim_{x\rightarrow c_2+0}R'(x)$. Moreover, the reinjection to the fixed point $x=1$ is
 determined by $\lim_{x\rightarrow c_1-0}R'(x)$ and $\lim_{x\rightarrow c_2-0}R'(x)$. Because of 
 $\lim_{x\rightarrow c_1+0}R'(x)=\lim_{x\rightarrow c_1-0}R'(x)$, $\lim_{x\rightarrow c_2+0}R'(x)=\lim_{x\rightarrow c_2-0}R'(x)$ 
 and $R'(x)>1$ on $(0,c_1]\cup [c_1,c_2] \cup [c_2,1)$, the way of a reinjection to $x=0$ and $x=1$ is the same. Since the behavior 
 near the indifferent fixed points is given by $S(x)-x\sim (x/c)^z(x\rightarrow 0)$ and $x-S(x)\sim \{(1-x)/(1-c)\}^z(x\rightarrow 1)$, 
 we can apply LTGA law to the reduced map $R(x)$. Then, the exponents $\alpha$ and $\beta$ are given by
\begin{equation}
\alpha=\frac{1}{z-1}\quad{\rm and}\quad
\beta = \left(\frac{c}{1-c}\right)^{\frac{z}{z-1}}.
\end{equation} 

First, we consider TAMSD, where the observation function $f_m(x)$ is the $L_{loc,\mu}^1(0,1)$ function with finite mean 
\cite{Akimoto2008}. By LTGA law, $f_m(0)=m^2$ and $f_m(1)=m^2$, TAMSD converges to $m^2$: 
$\overline{\delta x^2_m}=m^2$ for all $c$ and $z\geq 2$. We note that the ballistic behavior, $\overline{\delta x^2_m}=m^2$, 
is not due to a drift for $c=0.5$. Next, we consider the time-averaged drift (TAD) defined by the time average of $v_m(x)=T^m(x)-x$:
 \begin{equation}
\label{tad}
\overline{\delta x_m}  =\lim_{n\rightarrow\infty}\frac{1}{n}\sum_{k=0}^{n-1} v_m(x_k),
\end{equation}
which is also the $L_{loc,\mu}^1(0,1)$ function with finite mean. Figure 3 shows TADs obtained from different trajectories. 
By LTGA law, $v_m(0)=-m$ and $v_m(1)=m$, we have 
\begin{equation}
\Pr \left\{ \overline{\delta x_m}/m \leq t\right\}
\rightarrow
1-G_{\alpha,\beta}\left(\frac{t+1}{2}\right). 
\label{gasl.drift}
\end{equation}
This distributional limit theorem states that the time-averaged velocity, defined by $\overline{V} \equiv \overline{\delta x_m}/m$, 
under a bias is intrinsically random, {\it i.e.}, {\it distributional response}. Numerical simulations are in good agreement with the theory 
(see Fig. 4). The ensemble average of $\overline{V}$ is given by
\begin{equation}
V\equiv \langle \overline{V} \rangle_F = \frac{1-\beta}{1+\beta},
\label{response}
\end{equation}
where $\langle \cdot \rangle_F$ is the ensemble average under an external bias. Figure 4 shows the response of $V$ to a bias $c$.

{\it Generalized Einstein relation}.---The asymmetric parameter $c$ with $0<c<1$ is considered to be the probability of leftward walk 
if a persistent motion is terminated. Let $c=e^{-\frac{F}{kT}}/(e^{-\frac{F}{kT}}+e^{\frac{F}{kT}})$ be the leftward walk probability, where 
$F$ is an external force, $T$ is a temperature, and $k$ is the Boltzmann constant \cite{bouchaud90}. 
We consider a small bias, $c\cong 1/2 - F/kT$ for $F\rightarrow 0$. Expanding $\beta$ 
around $c=1/2$ and substituting it into $V$, we obtain the linear response of $V$ to a bias $F$:
\begin{equation}
V \sim -2\frac{z}{z-1} \frac{F}{kT}.
\end{equation}
Although TADs are intrinsically random, the TAMSDs are not random and grow as $m^2$ for $z\geq 2$. Therefore, we have the following 
generalized Einstein relation for superdiffusion:
\begin{equation}
\langle \overline{\delta x_m} \rangle_F =  -2\frac{z}{z-1}\frac{F}{kT}\sqrt{\overline{\delta x^2_m}}.
\label{ge}
\end{equation}

{\it Discussion}.---
We have found a distributional response in deterministic superdiffusion using the distributional limit theorem in infinite ergodic theory.  
In a recent study we obtained a generalized Einstein relation for single trajectories using Hopf's ergodic theorem \cite{Akimoto2011.arxiv}. 
However, this theorem does not work in deterministic superdiffusion because the observation function $v_m(x)$ is not 
an $L^1(\mu)$ function. Therefore, a distributional response is essential in superdiffusion. Moreover, we noted that the generalized Einstein 
relation (\ref{ge}) is different from that in anomalous diffusion \cite{bouchaud90}. We hope that our finding, {\it i.e.}, the distributional response, 
will be observed in experiments of random time-averaged observables.\par

The author thanks Tomoshige Miyaguchi for discussions. 
This work was partially supported by Grant-in-Aid for Young Scientists (B) (No. 22740262).




%

\begin{figure}
\includegraphics[height=.5\linewidth, angle=0]{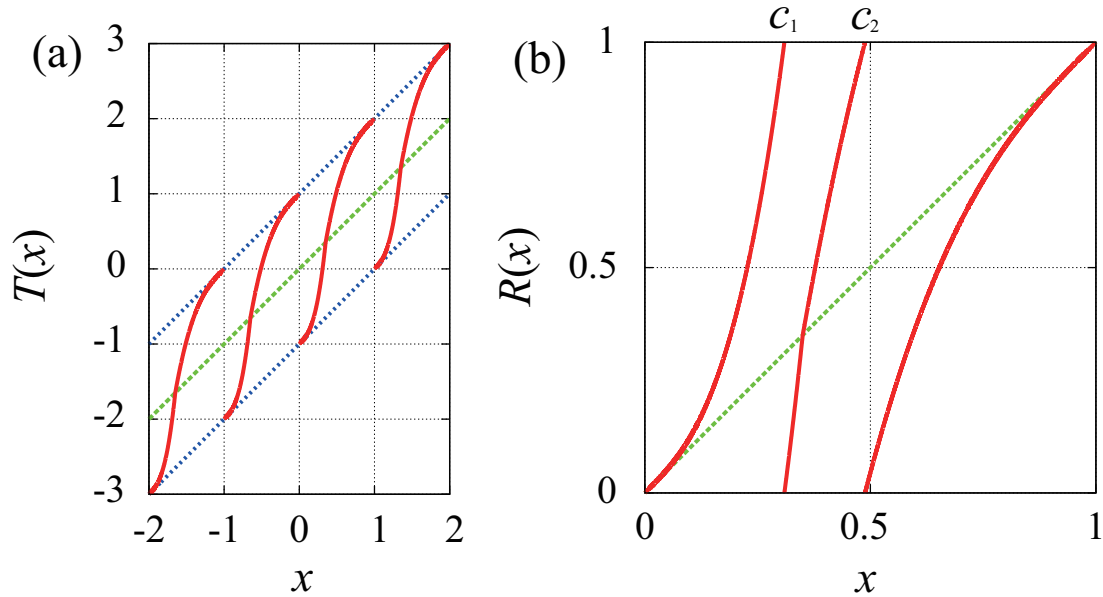}
\caption{ (Color online) (a) Asymmetric deterministic diffusion $T(x)$  $(z=3.0$ and $c=0.35$). (b) Its reduced map $R(x)$.}
\end{figure}

\begin{figure}
\includegraphics[height=.7\linewidth, angle=0]{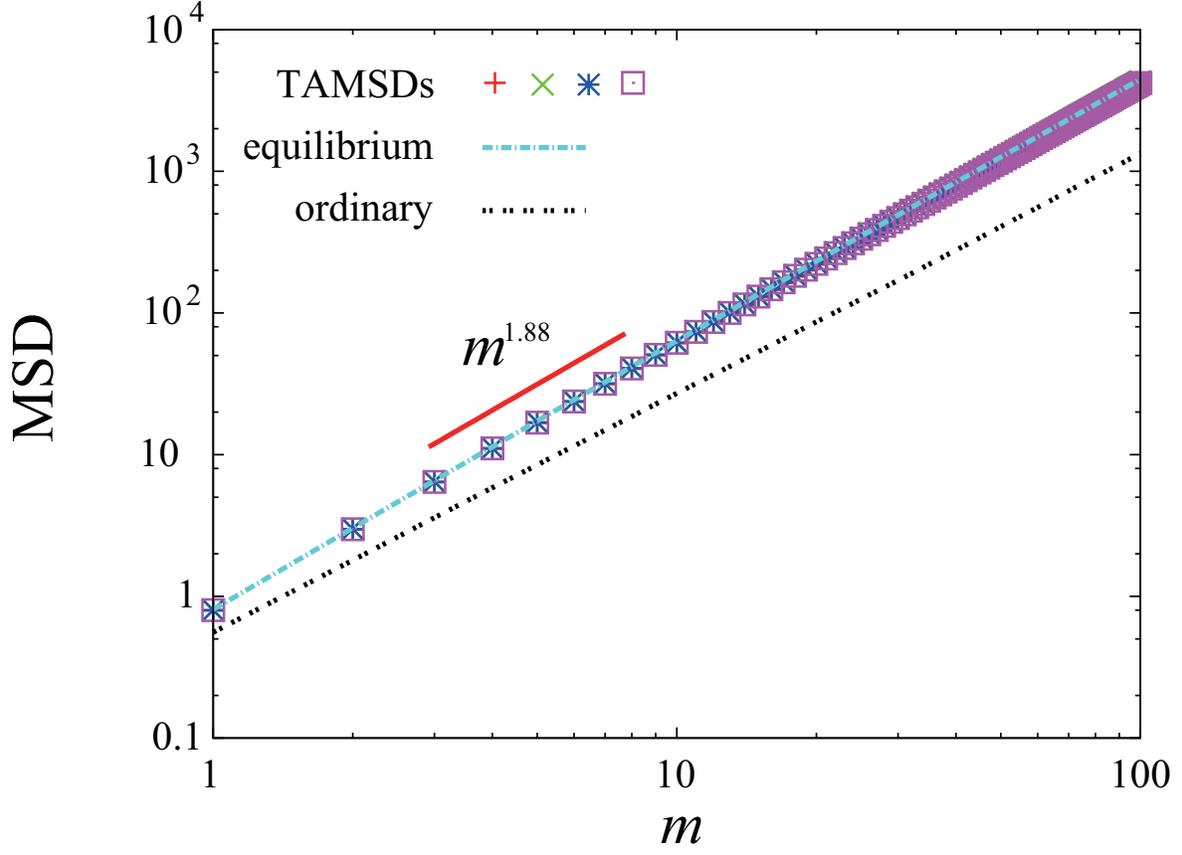}
\caption{ (Color online) Mean square displacements ($z=1.9$ and $c=0.5$). Different symbols are the TAMSDs calculated by 
different initial points. Dashed and dotted lines are the EAMSDs based on equilibrium and ordinary ensembles, respectively, 
where an ordinary ensemble is a uniform ensemble on $[0,1]$ and an equilibrium ensemble is the points after $10^6$ times 
iterations. The slope of the solid line is the theoretical exponent (\ref{msd}).}
\end{figure}

\begin{figure}
\includegraphics[height=.7\linewidth, angle=0]{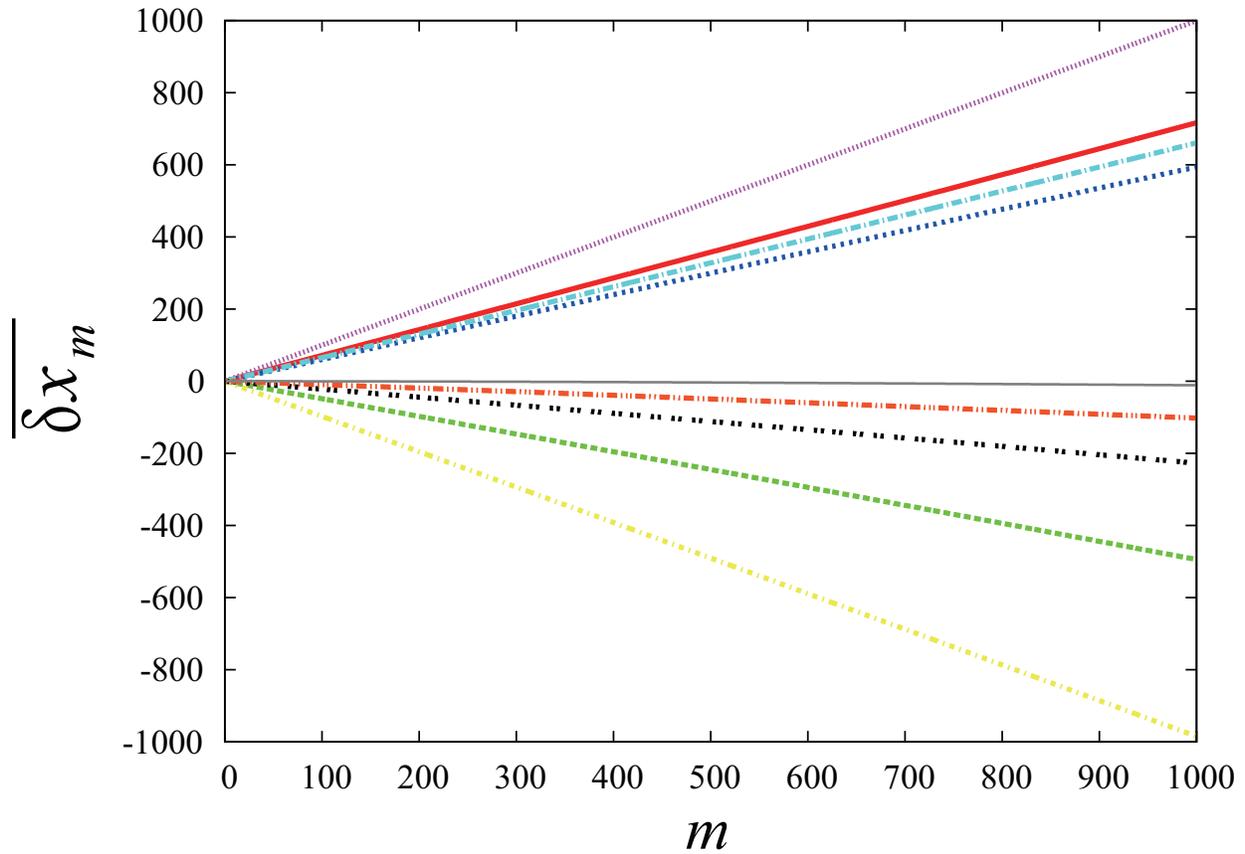}
\caption{ (Color online)  Time averaged drift $(z=3.0, c=0.45$, and $n=10^5$). TADs are calculated by different initial points.}
\end{figure}

\begin{figure}
\includegraphics[height=.7\linewidth, angle=0]{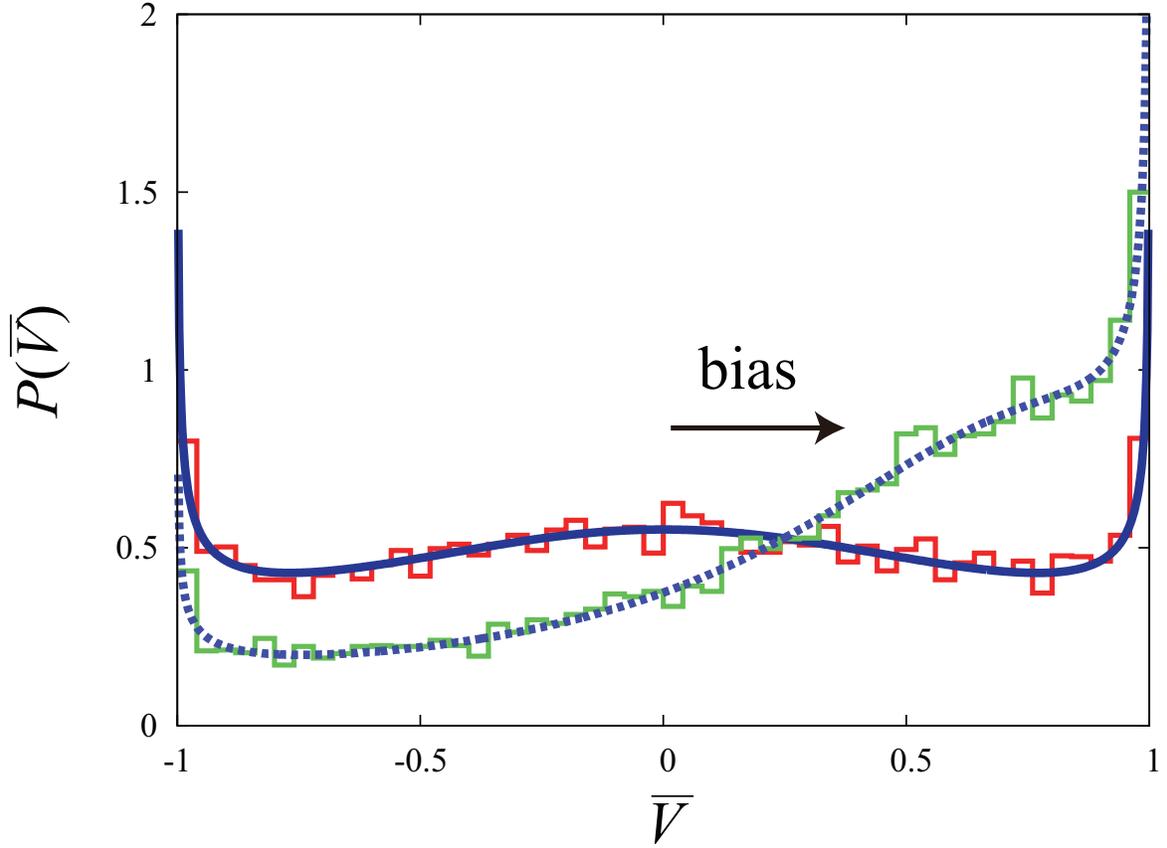}
\caption{ (Color online) Probability density function of $\overline{V}$ ($z=2.5$ and $n=10^7$). Numerical results are represented 
by the solid and dashed histograms with the theoretical PDFs (\ref{gasl.drift}) for $c=0.5$ and $c=0.4$, respectively. }
\end{figure}

\begin{figure}
\includegraphics[height=.7\linewidth, angle=0]{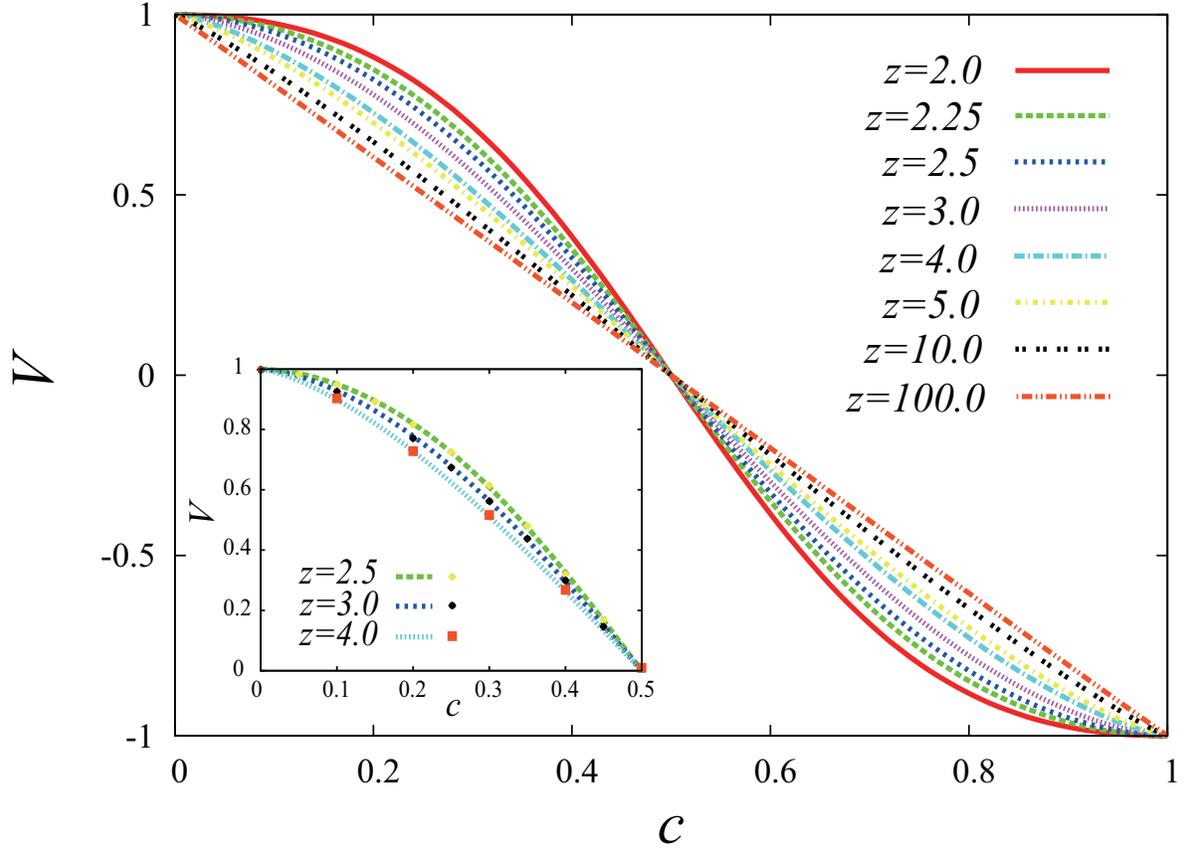}
\caption{ (Color online) Response to biases. Different curves are the response curves (\ref{response}) for different $z$. 
Symbols are the results of numerical simulations in the inset figure.}
\end{figure}

\end{document}